\documentclass{nature}

\usepackage[utf8]{inputenc}
\usepackage{graphicx}
\usepackage{epsfig}
\usepackage{hyperref}
\usepackage[T1]{fontenc}
\usepackage{amsmath}
\usepackage{amssymb}
\usepackage{braket}
\usepackage{color}
\usepackage[font=sf,small, labelfont=sf,small, textfont=sf,small]{caption}

\title{Time-frequency quantum process tomography of parametric down-conversion}

\author{Malte Avenhaus$^{1,2}$, Benjamin Brecht$^{2,\star}$, Kaisa Laiho$^{2,3}$, and Christine Silberhorn$^2$}

\begin{document}

\maketitle

\begin{affiliations}
	\item {Max Planck Institute for the Science of Light, G\"unther-Scharowsky Stra{\ss}e 1 - Building 24, 91058 Erlangen, Germany}
	\item {Integrated Quantum Optics, Applied Physics, University of Paderborn, Warburger Stra{\ss}e 100 33098, Paderborn, Germany}
	$^\star$e-mail:benjamin.brecht@uni-paderborn.de
	\item {University of Innsbruck, Institute for Experimental Physics/Photonics, Technikerstr. 25d, 6020 Innsbruck, Austria}
\end{affiliations}

\begin{abstract}
Parametric down-conversion (PDC) is the established standard for the practical generation of a multiplicity of quantum optical states. These include two-mode squeezed vacuum\cite{OuZY:1992wj}, heralded non-Gaussian states\cite{Lvovsky:2001tq,Ourjoumtsev:2006je,NeergaardNielsen:2006hl,Wakui:2007wz,Gerrits:2010iq} and entangled photon pairs\cite{Ou:1990bs,Kwiat:1995ck}. 
Detailed theoretical studies provide insight into the time-frequency (TF) structure of PDC, which are governed by the complex-valued joint spectral amplitude (JSA) function\cite{Grice:1997tk,Keller:1997hj,Grice:9999ud}. However in experiments, the TF structure of PDC is mostly characterised by \textit{intensity} measurements\cite{Wasilewski:2006td,Poh:2007gh,Avenhaus:2009vr,Chen:2009vo}that forbid access to the important phase of the JSA\cite{Brecht:2013fq}. 
In this paper, we present an amplitude-sensitive quantum process tomography technique that combines methods from ultrafast optics and classical three-wave mixing. Our approach facilitates a direct and phase-sensitive time-frequency tomography of PDC with high spectral resolution and excellent signal-to-noise ratio. 
This is important for all quantum optical applications, which rely on engineered parametric processes and base on minute knowledge of the quantum wave-function for the generation of tailored photonic quantum states. 
\end{abstract}
%
%


The accurate characterisation of quantum processes is essential for any system design, but at the same time poses a demanding task. %
The Hamiltonian associated with the process generates a unitary time evolution that is described by a complex-valued matrix. Thus, gaining complete knowledge on the process means to reveal both, the real and imaginary part of all matrix coefficients. This can be achieved by monitoring the action of the process on a chosen set of basis functions\cite{Poyatos:1997jh}. %
This procedure has been successfully implemented for phase-space tomography of a quantum process, deploying standard homodyne measurements on a basis set of coherent states\cite{Lobino:2008gg}. %
Note however that standard homodyne detection measures in the photon-number basis and largely neglects the TF properties of the process. The TF structure can only be recovered with large experimental effort, because many measurements and involved data processing techniques are required\cite{Lvovsky:2010fz,Polycarpou:2012kna,Roslund:2013cb}. This overhead renders a high-resolution retrieval of TF characteristics unfeasible. %

In this paper we follow a different path that was recently referred to as \textit{stimulated emission tomography}\cite{Liscidini:2013cx} and demonstrate that it is better adapted to the quick and reliable characterization of the complete TF properties of PDC. To this end, we focussed on waveguided PDC which provides a clean model system, since the spatial and spectral-temporal degrees of freedom are intrinsically decoupled. %

%
%
PDC is a three-wave mixing process, in which a bright pump field couples to the quantum mechanical vacuum such that during the propagation a pump photon decays into a pair of photons, usually labeled signal and idler. Formally, the unitary operation of this process may be written as %
\begin{equation}
	\hat{U}_\mathrm{PDC} = \exp\left[-\imath\left(\mathcal{B}\int d\omega_\mathrm{s}\,d\omega_\mathrm{i}\,f(\omega_\mathrm{s},\omega_\mathrm{i})\hat{a}^\dagger(\omega_\mathrm{s})\hat{b}^\dagger(\omega_\mathrm{i})+\mathrm{h.c.}\right)\right]
\end{equation}
where the constant $\mathcal{B}$ is a measure for the PDC gain. The function $f(\omega_\mathrm{s},\omega_\mathrm{i})$ is the complex-valued JSA, which contains the complete TF structure of the PDC and the $\hat{a}^\dagger(\omega_\mathrm{s})$ and $\hat{b}^\dagger(\omega_\mathrm{i})$ are standard creation operators generating signal and idler photons at frequencies $\omega_\mathrm{s}$ and $\omega_\mathrm{i}$. We can decompose the JSA function into two constituents, namely the pump envelope function $f_\mathrm{p}(\omega_\mathrm{s}+\omega_\mathrm{i})$ and the phasematching function $f_\mathrm{k}(\omega_\mathrm{s},\omega_\mathrm{i})$ \cite{Grice:9999ud}with %
\begin{equation}
	f(\omega_\mathrm{s},\omega_\mathrm{i}) =\mathcal{N}
	\underbrace{\exp\left[-\frac{(\omega_\mathrm{s}+\omega_\mathrm{i})^2}
	{\sigma_\mathrm{p}^2}\right]}_{=:f_\mathrm{p}(\omega_\mathrm{s}+
	\omega_\mathrm{i})}
	\cdot
	\underbrace{
	\mathrm{sinc}\left(\frac{\Delta k(\omega_\mathrm{s},\omega_\mathrm{i})}{2L}
	\right)
	\exp\left[\imath\frac{\Delta k(\omega_\mathrm{s},\omega_\mathrm{i})}
	{2L}\right]}_{=:f_\mathrm{k}(\omega_\mathrm{s},\omega_\mathrm{i})}.
\end{equation}
Here, $\sigma_\mathrm{p}$ denotes the spectral $\frac{1}{e}-$width of the PDC pump field, $\Delta k(\omega_\mathrm{s},\omega_\mathrm{i})$ is the phasemismatch between the pump, signal and idler and $L$ is the length of the nonlinear waveguide. The constant $\mathcal{N}$ is a normalisation constant that ensures $\int d\omega_\mathrm{s}\,d\omega_\mathrm{i}\,|f(\omega_\mathrm{s},\omega_\mathrm{i})|^2=1$. %

%
%
To retrieve the complete JSA function, we recall that PDC - as a three-wave mixing process - implements phase-sensitive amplification when seeding the signal and idler. This means that, depending on their phases relative to the pump, the seed fields are either amplified of depleted. %
We model the seed fields as delta-like distributions $\alpha(\omega)=\alpha e^{\imath\phi_\alpha}\delta(\omega-\omega_\alpha)$ and $\beta(\omega)=\beta e^{\imath\phi_\beta}\delta(\omega-\omega_\beta)$, where $|\alpha|^2$ and $|\beta|^2$ are the intensities of the seed fields, $\phi_\alpha$ and $\phi_\beta$ are their phases with respect to the PDC pump, and $\omega_\alpha$ and $\omega_\beta$ are their central frequencies, respectively. Amplification and depletion of the seed fields are hence governed by $\phi_\alpha+\phi_\beta$. Consequently, we find for the intensity of either of the seed fields after the PDC process (for details see the supplementary information): %
\begin{equation}
	\Delta I_{\alpha/\beta}\propto2|\alpha\beta f(\omega_\alpha,\omega_\beta)|\cos(\phi_\alpha+\phi_\beta+\phi_\mathrm{JSA}).
	\label{eq:intensity}
\end{equation}
Here, $\phi_\mathrm{JSA}$ is the phase contribution of the JSA function at the point $(\omega_\alpha,\omega_\beta)$, which for a standard PDC ranges between 0 and $\pi$. Measuring the intensity in one arm of the PDC with a linear photodiode grants direct access to both, the modulus and the phase of the JSA function, if the intensities and phases of the seed fields are known. Note that even without knowledge of the phases, our approach is superior to common characterisation methods, which retrieve only to the spectral intensity $|f(\omega_\mathrm{s},\omega_\mathrm{i})|^2$. There, an unfavorable signal-to-noise ratio masks important features of the JSA such as the sinc-function sidelobes. %
%
%

In our experiment shown in Fig. 1, we deployed a commercially available waveguide chip of periodically poled potassium titanyl phosphate (for more details see Methods). To realize different waveguide lengths $L$ without actually cutting the crystal, the waveguide strips feature different periodic poling lengths, hence different PDC interaction lengths. Prior to the measurements, we carefully characterized the properties of the different waveguides. We could retrieve the phasematching function $f_k(\omega_\mathrm{s},\omega_\mathrm{i})$, which was oriented along an angle of $-35^\circ$ with respect to the $\omega_\mathrm{s}$-axis and had a spectral width ranging from $0.5\,$nm to $2.5\,$nm, depending on the waveguide length. %
%
%

\begin{figure}
	\centering
	\includegraphics[width=170mm]{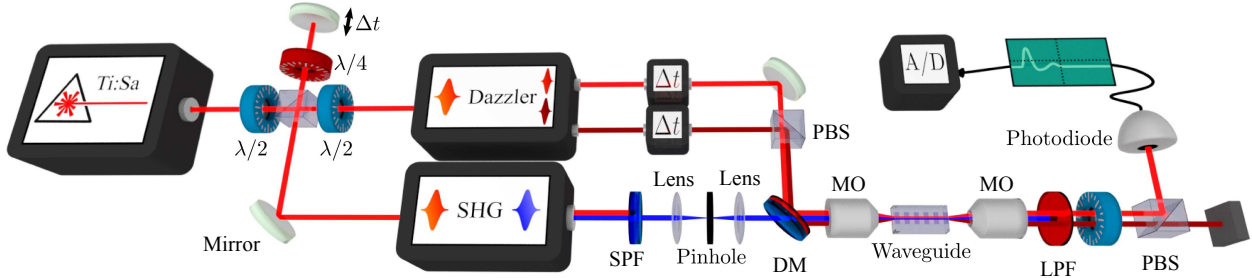}
	\caption{\textbf{Experimental setup} An ultrafast cavity-dumped Ti:Sapph oscillator (KMLabs Cascade-5) was used to generate both, the PDC pump and the seed fields. After a variable temporal delay for pulse synchronization, a small portion of the Ti:Sapph radiation was sent to the pulse shaper (Fastlite Dazzler) which was used to independently shape the two seed fields in orthogonal polarizations. Each seed field was again temporally delayed and then the fields were recombined on a polarizing beamsplitter (PBS). The main part of the laser light was sent to a second-harmonic generation (SHG) with a successive short pass filter (SPF), to suppress the remaining fundamental radiation. Then, a spatial mode cleaner consisting of two lenses and a pinhole was deployed to optimize the SHG, which served as pump for the PDC. Seed fields and pump pulses were combined on a dichroic mirror (DM) and coupled to the nonlinear waveguide with a $20\times$ microscope objective (MO). Behind the waveguide, a long-pass filter (LPF) and another PBS were used to separate the pump and the modified seed fields from each other. One seed was detected with a \textit{linear} avalanche photo diode and its electric response signal was digitized with a fast A/D transient recorder.}
  \label{fig:Setup} 
\end{figure}

To generate and tailor the seed fields, we used a programmable accousto-optic pulse shaper, Dazzler, from Fastlite, with a spectral resolution of $0.1\,$nm and a minimal bandwidth of $0.3\,$nm. Here, acoustic and optical pulses travel in quasi-collinear configuration. Since the acoustic velocity is much slower than their optical counterpart, a burst of 100 optical pulses is shaped while the acoustic pulse traverses the Dazzler. The dispersion properties of the Dazzler inevitably lead to a change of $\phi_\alpha+\phi_\beta$ between successive pulses. For more details see the Methods section. %

We monitored the intensity $I_\alpha$ in one arm of the PDC with a Hamamatsu S5343 avalanche photo diode (APD) operated in linear amplification regime. Typical measurement traces are shown in Fig. 2. Here, we plot the APD signal against the number of the optical pulse in the burst for different sampling points $(\omega_\mathrm{s},\omega_\mathrm{i})$. We observe alternating amplification and de-amplification of the signal, in accordance with equation (\ref{eq:intensity}), where the period of the oscillation is defined by the dispersion of the pulse shaper. The amplitude of the oscillations provides direct access to the modulus of the JSA function, whereas its period does not play a role for the reconstruction. %

The yellow trace in Fig. 2 corresponds to a frequency combination $(\omega_\alpha,\omega_\beta)$, for which the dispersion of the Dazzler leads to a phase change of $\phi_\alpha+\phi_\beta=2n\pi$ with $n$ being a (large) integer. Consequently, we observe no oscillations in $I_\alpha$. However, these traces are extremely valuable for the reconstruction of the actual phase of the JSA $\phi_\mathrm{JSA}$, as described below. %

%
%

\begin{figure}
	\centering
	\includegraphics[width=88mm]{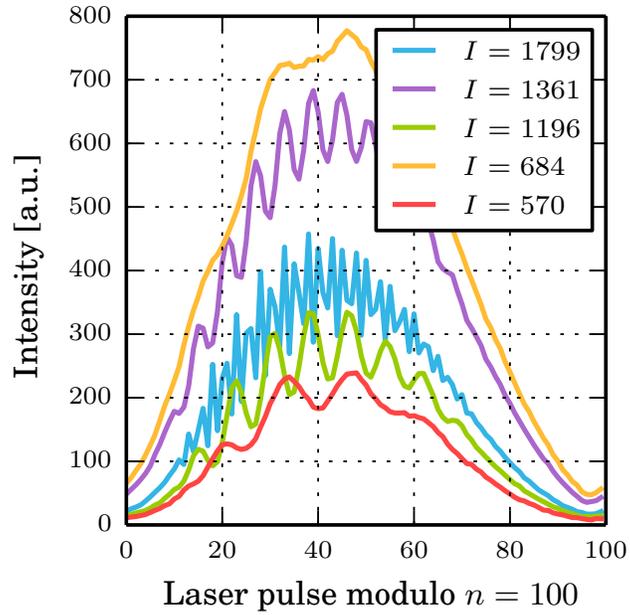}
    \caption{\textbf{Measurement trace} Typical measurement traces for sampling different points of the JSA, as recorded with the APD. The numbers in the legend refer to internal encodings of $\omega_\mathrm{s}$-$\omega_\mathrm{i}$ combinations. Alternating amplification and de-amplification between successive pulses of a burst are visible for all traces, except for the yellow one. The period of the oscillations is defined by the dispersion of the pulse shaper. Note that for the reconstruction of the JSA, only the oscillation contrast is required. }
	\label{fig:PulseIntensity}
\end{figure}

\begin{figure}
	\centering
	\includegraphics[width=170mm]{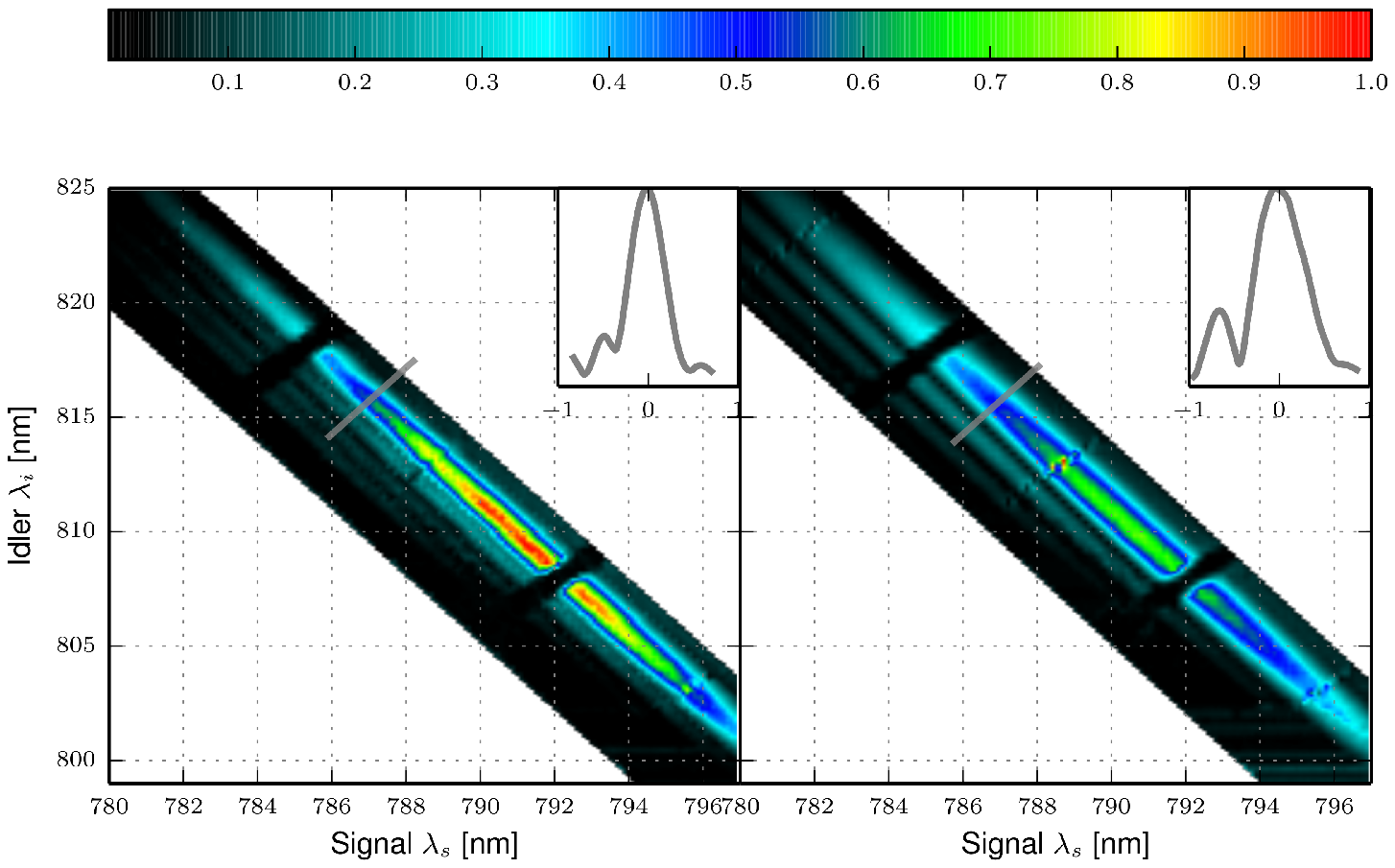}
    \caption{\textbf{Modulus of the JSA} Our method facilitates the characterization of the modulus of the JSA $f(\omega_\mathrm{s},\omega_\mathrm{i})$ with exceptionally high spectral resolution. The sidepeaks of the sinc-shaped phasematching are clearly discernible both for a $2.5\,$mm long waveguide (left) and a $1.0\,$mm long waveguide (right). The phasematching function for the shorter waveguide is considerably broader as indicated by the insets in the upper right corners which show a cut through $f(\omega_\mathrm{s},\omega_\mathrm{i})$. In addition, the peak height of the JSA is lower for the shorter waveguide. This last result highlights that our characterization method also allows for quantitative statements regarding the expected efficiency of the PDC.}
	\label{fig:JSA}
\end{figure}

We scanned the JSA function for different waveguides on our sample. In Fig. 3, we plot the measured JSA functions for waveguides with interaction lengths of $2.5\,$mm and $1.0\,$mm, respectively. The sidelobes of the sinc-shaped phasematching function can be clearly seen in the measurement. This highlights the outstanding resolution of our approach, which in our case is only limited by the resolution of the pulse shaper. The black regions in Fig. 3 correspond to the aforementioned period constellation where no oscillations in $I_\alpha$ are observed. %

Finally, in Fig. 4, we present the mapped phase of the JSA. For this purpose, we recorded the intensity at one specific pulse number when scanning the seed frequencies $\omega_\alpha$ and $\omega_\beta$ across one of the black regions in Fig. 3. In this configuration, $\phi_\alpha+\phi_\beta$ is constant, hence any changes in the measured intensity are based on a change in $\phi_\mathrm{JSA}$. In the top plot, we evaluate the measurements only in terms of amplification and de-amplification of the seed and obtain a stepwise alternation between the two possible configurations. By combining this result with the measurements from Fig. 3, we finally retrieve the complete, complex JSA function of the PDC under investigation and show its real part in Fig. 4 (bottom plot).%

%
%

\begin{figure}
	\centering
	\includegraphics[width=88mm]{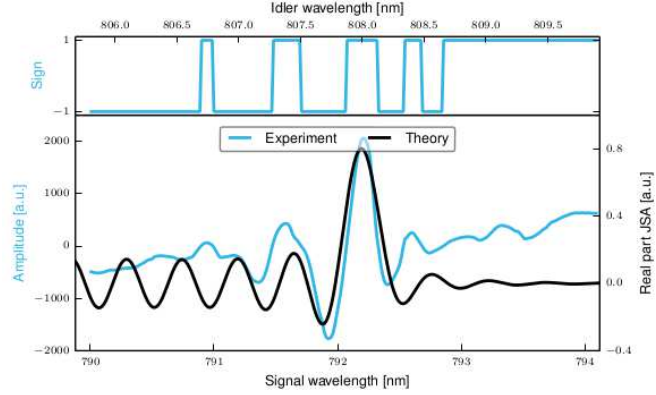}
    \caption{\textbf{Phase of the JSA} The phase of the JSA can be retrieved when evaluating the amplification / de-amplification of the signal seed field, while scanning the seed frequencies along one of the black regions from Fig. 3. This measurement yields a binary behavior, as can be seen in the top plot. Without loss of generality, we identify a total phase of $\phi_\mathrm{JSA}=0$ with the main peak of the JSA. Consequently, $\phi_\mathrm{JSA}$ jumps between 0 and $\pi$. Combining these results with the recovered amplitude distribution finally yields the complex JSA, as shown in the bottom plot.}
	\label{fig:JSA-Phase}
\end{figure}

In this paper we have introduced a novel method for a fast TF quantum process tomography of PDC. By phase-sensitive seeding of the PDC with two coherent fields, we demonstrated the direct measurement of the complex-valued JSA function which characterizes the complete PDC TF structure. In contrast to quantum process tomography protocols based on standard homodyne detection, our approach is highly adapted to measuring the spectral-temporal characteristics of a PDC and can be implemented with off-the-shelf components. Hence, we expect this characterization technique to have an impact on many quantum optical applications relying on engineered parametric processes for the tailored generation of exotic quantum states of light. 
%
%


\begin{methods}
The cavity-dumped Ti:Sapph laser had a spectral bandwidth of $50\,$nm, and was driven at a repetion frequency of $2\,$MHz. The part of the laser light that formed the seeds was sent to the Dazzler with a polarization of $45^\circ$, which facilitated independent shaping of the H and V component. The remaining laser light was sent through the SHG, which we realized in a $3\,$mm long $\mathrm{BiB}_3\mathrm{O}_6$ crystal. The SHG with a FWHM of around $1.7\,$nm was tuneable from $396\,$nm to $401\,$nm by adjusting the tilt and provided up to $1.5\,$mW of cw-equivalent power. The asymmetric spatial mode of the SHG light was subsequently cleaned with a $20\,\mu$m diameter pin-hole before being coupled into the KTP waveguide from AdvR, which featured a cross-section of $4.3\,\mu$m$\times4.3\,\mu$m and a poling period of roughly $7.5\,\mu$m to provide type II phasematching. A DG645 digital delay generator from Stanford instruments was used to lock the Dazzler shaping frequency to the master clock of the Ti:Sapph laser, re-triggering the pulse shaping process every 100 optical pulses. 

The particular internal setup of the Dazzler enforces different optical path lengths for the two seeds. Let us consider a field that enters the Dazzler and is oriented along the fast axis of the birefringent Dazzler crystal. At a specific point in the crystal, the shaped light is diffracted by the travelling acoustic shaping wave to the slow axis of the Dazzler crystal. Hence, the total optical path of the light is described by $\Delta x_1= n_\mathrm{fast}x+n_\mathrm{slow}(L-x)$, where $x$ marks the position of light diffraction in the crystal of length $L$, and where $n_\mathrm{fast}$ and $n_\mathrm{slow}$ are the refractive indices of the fast and slow crystal axis, respectively. When the next optical pulse enters the Dazzler crystal, the traveling acoustic wave has propagated further through the crystal by $dx$. Hence, the new optical pulse experiences an optical path length given by $\Delta x_2 = n_\mathrm{fast}(x+dx)+n_\mathrm{slow}(L-x-dx)$. This consequently results in an optical pathlength difference $\Delta x_1-\Delta x_2$ between successive pulses and hence in an associated phase-difference $\phi_1-\phi_2$. It is for this reason that the Dazzler intrinsically implements a phase-scan between successive pulses in the shaped burst which depends on the precise knowledge of the dispersion on the one hand and the crystal length on the other hand. 

Finally, for measuring the JSA function we deployed seed fields with the minimum width of $0.3\,$nm, limited by the Dazzler's bandwidth. We scanned the JSA function on a $0.1\,$nm wavelength grid and recorded for each point the intensity. The electronic APD signal was first amplified with a $20\,$dB amplifier from FEMTO with a bandwidth of $200\,$MHz and subsequently digitized with a fast 12 bit A/D transient recorder (M2i.3024-exp, Spectrum) with 100MS/s. From a Fourier analysis of the oscillations we could retrieve the modulus of the JSA, where we note again that this value only depends on the amplitude of the oscillations but not on their period, which was determined by the Dazzler's dispersion.  
\end{methods}



\bibliography{text}

\begin{addendum}
	\item The authors acknowledge outstanding support from Nicolas Forget (Fastlite) on operating the Dazzler with two polarizations simultaneously.
	\item[Author Contributions] M.A. and C.S. conceived and designed the experiment. All authors contributes substantially to the manuscript.
	\item[Competing Interests] The authors declare that they have no competing financial interests.
	\item[Correspondence] Correspondence and requests for materials should be addressed to Benjamin Brecht (email:benjamin.brecht@upb.de)
\end{addendum}

\end{document}